\documentclass[useAMS,usenatbib]{mn2e}

\usepackage{verbatim} 
\usepackage{natbib} 
\usepackage{amsmath} 
\usepackage{amsbsy}
\usepackage{amssymb}
\usepackage{mathrsfs} 
\usepackage{lscape} 
\usepackage{graphicx}
\usepackage{epstopdf}
\usepackage{deluxetable} 
\usepackage{fixltx2e} 

\title[FeLoBALs \& QSO Feedback]{A Physical Model of FeLoBALs: Implications for Quasar Feedback}

\author[Faucher-Gigu\`ere, Quataert, \& Murray]{Claude-Andr\'e Faucher-Gigu\`ere\thanks{Miller Fellow;
cgiguere@berkeley.edu}$^{1}$, Eliot Quataert$^{1}$, Norman Murray$^{2,3}$ \vspace*{6pt}  \\ 
$^{1}$Department of Astronomy and Theoretical Astrophysics Center, University of California, Berkeley, CA 94720-3411, USA.\\
$^{2}$Canadian Institute for Theoretical Astrophysics, 60 St. George Street, University of Toronto, ON M5S 3H8, Canada.\\
$^{3}$Canada Research Chair in Astrophysics
}

\begin{document}
\maketitle


\begin{abstract}
Photoionization modeling of the low-ionization broad absorption lines of certain quasars, known as FeLoBALs, has recently revealed the number density of the wind absorbers and their distance from the central supermassive black hole. 
From these, the feedback efficiency of the quasars can in principle be derived. 
The implied properties of the FeLoBALs are, however, surprising, with the thickness of the absorbers relative to their distance from the black hole, $\Delta R/R$, as small as $\sim10^{-5}$. 
Such absorbers are unlikely to survive the journey from the supermassive black hole to their inferred location. 
We show that the observed FeLoBAL properties are readily explained if they are formed \emph{in situ} in radiative shocks produced when a quasar blast wave impacts a moderately dense interstellar clump along the line of sight. 
This physical picture differs significantly from the thin shell approximation often assumed, and implies outflow rates, kinetic luminosities and momentum fluxes that differ correspondingly, in some cases at the order of magnitude level.   Using the radiative shock model, we estimate the ratio of the outflow kinetic luminosity to bolometric luminosity for three bright FeLoBAL quasars in the literature. 
We find $\dot{E}_{\rm k}/L_{\rm bol}\approx 2-5\%$ (and corresponding momentum fluxes $\dot{P} \approx2-15 \, L_{\rm bol}/c$), similar to what is adopted in models reproducing the $M-\sigma$ relation.  These outflow properties are also comparable to those recently inferred for molecular outflows in local ultra-luminous infrared galaxies,  suggesting that active galactic nuclei are capable of driving such outflows. 
\end{abstract}


\begin{table*}
\centering
\caption{Sample of FeLoBALs and their Inferred Properties\label{FeLoBAL properties table}}
\begin{tabular}{|ccccccccccccc|}
\hline\hline
QSO                               &  $z$ & $L_{\rm bol}$\tablenotemark{1}  & $M_{\rm BH}$\tablenotemark{2} & $\log{N_{\rm H}}$ & $\log{n_{\rm e}}$ & $\log{U_{\rm H}}$\tablenotemark{3} & $T$ & $R$\tablenotemark{4} & $\Delta R$\tablenotemark{5} & $v$   \\ 
                                       &          & erg s$^{-1}$   & M$_{\odot}$ & cm$^{-2}$ & cm$^{-3}$  &  & K & kpc  & pc & km s$^{-1}$ \\ 
\hline
SDSS J0838+2955\tablenotemark{a} & 2.043 & $10^{47.5}$ & $3\times10^{9}$ & $20.8$ & $3.8$ & $-1.9$ & $\approx 10^{4}$ & $3.3$ & $0.04$ & 4,900 \\ 
SDSS J0318-0600\tablenotemark{b} & 1.967 & $10^{47.7}$ & $4\times10^{9}$ & $20.1$ & $3.3$ & $-3.0$ & $\approx 10^{4}$ & 5.5 & $0.02$ & 4,200 \\ 
QSO 2359-1241\tablenotemark{c} & 0.858 & $10^{46.7}$ & $4\times10^{8}$ & $20.6$ &  $4.4$ & $-2.4$ & $\approx 10^{4}$ & 1.3  & 0.005 & 1,400 \\ 
\hline
\tablenotetext{1}{Bolometric luminosity of the quasar.
}
\tablenotetext{2}{SMBH mass, assuming that the bolometric luminosity equals the Eddington luminosity, i.e. $M_{\rm BH}\equiv 10^{8}~{\rm M_{\odot}} (L_{\rm bol} / 1.3\times10^{46}~{\rm erg~s^{-1}})$.}
\tablenotetext{3}{Ionization parameter $U_{\rm H} \equiv \Phi_{\rm H} / c n_{\rm H}$, where $\Phi_{\rm H}$ is the hydrogen ionizing photon flux.}
\tablenotetext{4}{Distance of the absorber from the central source.}
\tablenotetext{5}{Thickness of the absorber, defined as $\Delta R \equiv 1.2 N_{\rm H}/n_{e}$, assuming that the system is of solar composition and fully ionized. 
}
\tablenotetext{a}{Component c \citep[][]{2009ApJ...706..525M}. $^{b}$Component i \citep[][]{2010ApJ...709..611D}. $^{c}$Component e \citep[][]{2008ApJ...688..108K, 2010ApJ...713...25B, 2010IAUS..267..350A}.}
\end{tabular}
\end{table*}

\section{INTRODUCTION}
The broad absorption lines (BAL) seen in up to 40\% of quasars \citep[e.g.,][]{2006ApJS..165....1T, 2008ApJ...672..108D}, and blueshifted by up to $\sim 0.1c$ relative to systemic, are clear signatures of outflows driven by active galactic nuclei (AGN). 
Such outflows are of particular interest in the context of the measured correlations between galactic-scale properties and those of the supermassive black holes (SMBH) they host, such as the $M-\sigma$ relation \citep[e.g.,][]{2000ApJ...539L...9F, 2000ApJ...539L..13G, 2002ApJ...574..740T}.  
Indeed, it has been argued that feedback by accreting SMBHs could regulate their growth and produce these relationships \citep[e.g.,][]{1998A&A...331L...1S, 2003ApJ...595..614W, 2005Natur.433..604D}. 
While these arguments are broadly compelling, it remains an open question how the energy and/or momentum released by the SMBH couples to the surrounding galaxy, quenches black hole accretion, and possibly truncates star formation. 
Observational constraints on the feedback provided by AGN are therefore extremely valuable, but have been scarce so far. 

Progress on this front has recently been achieved through the detailed study of a subset of BAL systems, for which photoionization modeling has allowed inference of their column density $N_{\rm H}$ and radius $R$ from the SMBH, in addition to their velocity $v$ \citep[e.g.,][]{2009ApJ...706..525M, 2010ApJ...709..611D, 2010ApJ...713...25B}. 
Because these systems show absorption from low-ionization states of iron, specifically FeII, they are known as FeLoBALs.\footnote{HeI, SiII, and NiII ions also provide powerful diagnostics, when present.} 
Assuming a thin shell geometry, these measurements have been interpreted as providing estimates of the mass outflow rates, $\dot{M}_{\rm out}$, and kinetic luminosities, $\dot{E}_{\rm k}$, of the quasar outflows. 
These studies of a handful of FeLoBALs indicate that in some cases the kinetic luminosity of the AGN outflow can exceed $f=1\%$ of the bolometric luminosity \citep[e.g.,][]{2009ApJ...706..525M}. Such values are not far from the value $f=5$\% thought to be sufficient to explain the normalization of the $M-\sigma$ relation (e.g., Di Matteo et al. 2005\nocite{2005Natur.433..604D}; see also DeBhur et al. 2011\nocite{2011MNRAS.412.1341D}).

The physical properties of the FeLoBALs from which the kinetic luminosities and mass outflow rates are estimated (Table \ref{FeLoBAL properties table}) are surprising and differ significantly from the interpretation of other BAL observations. 
In particular, whereas classical CIV BALs are inferred to be located at distances $R\lesssim1$ pc from the SMBH \citep[e.g.,][]{2011MNRAS.413..908C}, the FeLoBALs in Table \ref{FeLoBAL properties table} have $R\gtrsim1$ kpc. 
Furthermore, the small inferred FeLoBAL sizes $\Delta R\sim0.01$ pc imply large ratios $R/\Delta R\sim10^{5}$ and, as we will show, lifetimes $\sim10^{-2}$ of the time it would take such absorbers to travel from the SMBH to their inferred locations, $t_{\rm flow}\approx R/v$. 
To determine the reliability of these observational estimates of the efficiency of AGN feedback, it is important to have a more detailed physical picture of the FeLoBAL absorbers. 
Such a model is in particular necessary to determine how global parameters of the outflow can be derived from the absorption measurements. 

In this paper, we argue that the FeLoBALs must be formed \emph{in situ} at large radii (\S \ref{created in situ}). 
We show that the interaction of a quasar blast wave with a moderately dense interstellar clump along the line of sight can result in radiative shocks, and that cool material entrained by the hot ambient flow can explain the properties of the observed FeLoBALs (\S \ref{radiative shocks}). 
This physical picture has important implications for the kinetic luminosities of the quasar outflows probed by low-ionization absorption lines (\S \ref{inferring feedback}). 
We conclude in \S \ref{discussion} by discussing our results in the context of models of the co-evolution of black holes and galaxies. 

Before proceeding, a clarification of the terminology adopted here is in order. 
The different classes of broad quasar absorbers are usually defined in terms of their observational spectroscopic properties. 
While these are indicative of their physical nature, there need not always be a one-to-one correspondence. 
In particular, only a handful of FeLoBAL absorbers to date have been studied in sufficient detail to extract their physical properties. 
In this work, we assume that the systems in Table \ref{FeLoBAL properties table} are representative of FeLoBALs in general. 
Strictly speaking, it is conceivable that some FeLoBALs are physically distinct and formed differently than we propose \citep[e.g.,][]{2011MNRAS.411.2653H}. 
Furthermore, some members of the broader class of low-ionization broad absorbers (LoBALs), or even some high-ionization BALs (HiBALs), may form as we describe here.   More detailed observations and photoionization analyses will be needed to determine exactly how widely applicable the model presented here is.

Our model is based on the crushing of an interstellar cloud by the forward shock driven into the ambient medium by the supersonic quasar outflow. A review of the important cloud crushing physics can be found in \cite{1994ApJ...420..213K}. A key element is that the shocks driven into the over-dense cloud are much weaker than the forward shock in the ambient gas, which allows them to cool and dust to survive (\S \ref{radiative shocks}). A third shock involved is the reverse shock into the quasar wind itself.

\begin{figure*}
\begin{center} 
\includegraphics[width=1.0\textwidth]{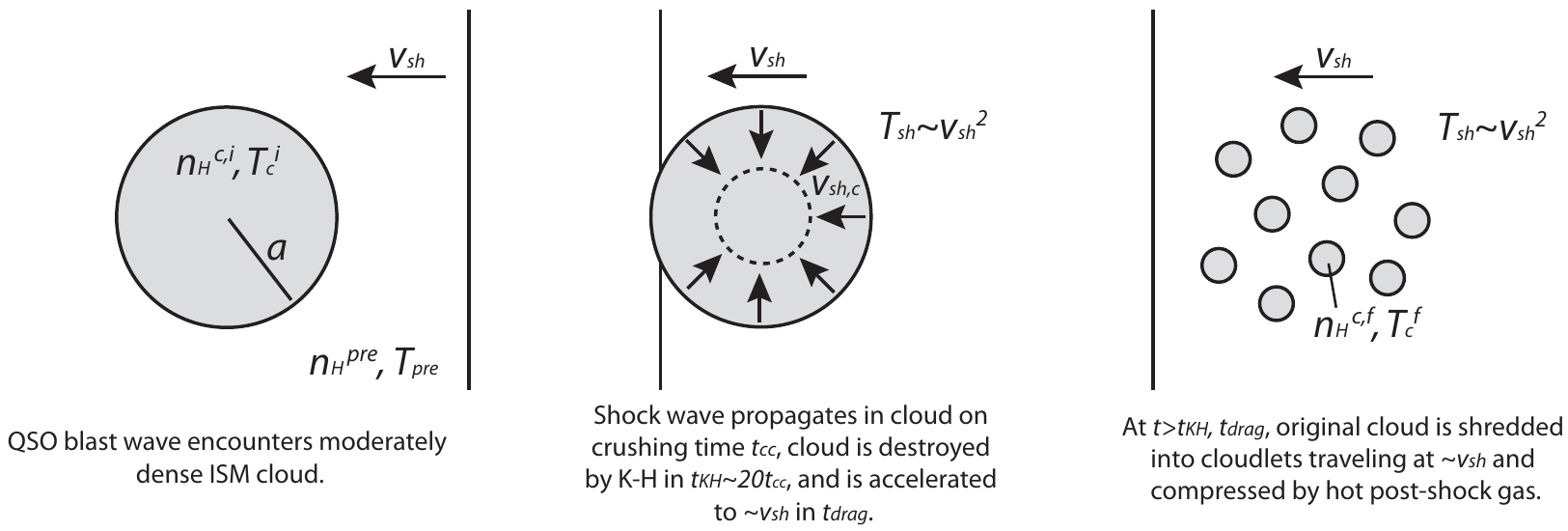}
\end{center}
\caption[]{Schematic illustration of the formation of FeLoBALs in radiative shocks.}
\label{cartoon} 
\end{figure*}

\section{MODEL}
\label{model}

\subsection{FeLoBALs are Created \emph{In Situ}}
\label{created in situ}
Quasar wind ejecta with velocity $v$ take a flow time  
\begin{equation}
t_{\rm flow} \approx \frac{R}{v} \approx 3\times 10^{5}{\rm~yr}\left( \frac{R}{\rm 3~kpc} \right) \left( \frac{v}{\rm 10,000~km~s^{-1}} \right)^{-1}
\end{equation}
to travel out to a radius $R$.\footnote{If the ejecta decelerate by a factor of a few along the way, the required time scale is correspondingly shorter, but this does not significantly affect our arguments.}  
As the ejecta impact the interstellar medium (ISM) of the galaxy, the gas behind the shock is heated to a temperature
\begin{equation}
\label{Tsh}
T_{\rm sh}=\frac{3 \mu}{16k}m_{\rm p} v_{\rm sh}^{2}\approx 1.3\times10^{9}~{\rm K} \left( \frac{4v/3}{\rm 10,000~km~s^{-1}} \right)^{2},
\end{equation}
where $\mu$ is the mean molecular weight (assuming $\mu=0.61$ for a fully ionized gas of solar composition) and $k$ is Boltzmann's constant.
In this expression, we have assumed $v=3v_{\rm sh}/4$, as appropriate for the post-shock material for a strong shock. 
For comparison, the sound crossing time of a cloud of size $\Delta R$ is
\begin{equation}
\label{sound crossing time}
t_{\rm s} \approx \frac{\Delta R}{c_{s}}  \approx 640~{\rm yr} \left( \frac{\Delta R}{\rm 0.01~pc} \right) \left( \frac{T}{\rm 10^{4}~K} \right)^{-0.5},
\end{equation}
so that it would have ample time to adjust its size. 
If the cloud has temperature $T_{\rm BAL}$, number density $n_{\rm H}^{\rm BAL}$, and is pressure confined by a medium at temperature $T_{\rm sh}$, that medium must have density
\begin{align}
\label{nH hot}
&n_{\rm H}^{\rm hot} = \frac{T_{\rm BAL}}{T_{\rm sh}} n_{\rm H}^{\rm BAL}  \\ \notag
& \approx 0.08~{\rm cm^{-3}} \left( \frac{n_{\rm H}^{\rm BAL}}{\rm 10^{4}~cm^{-3}} \right) \left( \frac{T_{\rm BAL}}{\rm 10^{4}~K} \right) \left( \frac{T_{\rm sh}}{\rm 1.3\times10^{9}~K} \right)^{-1}.
\end{align}

Non-gravitationally bound cool clouds (such as those in Table \ref{FeLoBAL properties table}) moving in a hot medium are disrupted by the Kelvin-Helmholtz instability on a time scale
\begin{equation}
\label{KH time scale}
t_{\rm KH} \approx \kappa \left( \frac{n_{\rm H}^{\rm BAL}}{n_{\rm H}^{\rm hot}} \right)^{1/2} \frac{\Delta R}{v},
\end{equation}
where the velocity is relative to the hot medium, and the dimensionless factor can have values up to $\kappa \approx 10$ when cooling is efficient \citep[e.g.,][]{2002A&A...395L..13M, 2009ApJ...703..330C}, but not much more. 
For the fiducial parameters in equations (\ref{sound crossing time}) and (\ref{nH hot}), assuming that $v=c_{\rm s}(T_{\rm sh})$, equation (\ref{KH time scale}) yields $t_{\rm KH}\approx 630 \kappa$ yr. 
Since $t_{\rm KH} \ll t_{\rm flow}$ for the FeLoBALs in Table \ref{FeLoBAL properties table}, these absorption systems would have been completely destroyed along the way if they had been produced in the immediate vicinity of the SMBH, as is inferred to be the case for many classical, high-ionization BAL absorbers. 
While the Kelvin-Helmholtz instability can in principle be suppressed by magnetic fields, the Rayleigh-Taylor instability generically cannot \citep[][]{2007ApJ...671.1726S}, and would destroy the clouds on a comparable time scale. 
Furthermore, thermal conduction from the hot medium would evaporate the cool clouds on a time scale 
\begin{align}
t_{\rm evap} \approx 6\times10^{3}~{\rm yr}~
\left(
\frac{\Delta R}{\rm 0.01~pc}
\right)^{7/6}
& \left( 
\frac{n_{\rm H}^{\rm BAL}}{\rm 10^{4}~cm^{-3}}
\right) \\ \notag
& \times \left(
\frac{n_{\rm H}^{\rm BAL} T_{\rm BAL}}{\rm 10^{8}~K~cm^{-3}}
\right)^{-5/6} 
\end{align}
\citep[saturated heat flux regime;][]{1977ApJ...211..135C} regardless.

Unlike the BALs directly accelerated near SMBHs, for example via disk winds \citep[e.g.,][]{1995ApJ...451..498M}, the FeLoBALs in Table \ref{FeLoBAL properties table} from which mass outflow rates and kinetic luminosities have been estimated therefore very likely formed at the large radii where they are observed.  

\subsection{Formation in Radiative Shocks}
\label{radiative shocks}
At velocities $\gtrsim1,000$ km s$^{-1}$, the post-shock gas cools predominantly by free-free emission, losing its energy at constant pressure on a time scale 
\begin{equation}
t_{\rm cool} \approx 10^{8}~{\rm yr} \left( \frac{T_{\rm sh}}{10^{9}~{\rm K}} \right)^{1/2} \left( \frac{n_{\rm H}^{\rm pre}}{\rm 1~cm^{-3}} \right)^{-1},
\end{equation}
where we assume that the shock is strong, so that the post-shock density is 4 times the pre-shock density $n_{\rm H}^{\rm pre}$. 
A limit on the average number density swept up by the quasar outflow can be derived using momentum conservation.

Over a time scale $t_{\rm flow}$, the quasar injects a momentum 
\begin{equation}
P = f_{\rm w} \frac{L_{\rm bol}}{c} t_{\rm flow}
\end{equation}
into a wind. 
The factor $f_{\rm w}$ can be larger than unity, for example in  a confined shocked wind doing work on surrounding gas or if there are multiple photon scatters in an optically thick medium \citep[e.g.,][]{2005ApJ...618..569M}; alternatively, the mass inflow and outflow rates can in principle exceed the Eddington rate.  
If $\bar{\rho}_{\rm pre}(<R)$ is the average gas mass density enclosed within a radius $R$, where the velocity of the outflow is $v$, 
\begin{equation}
P \approx \frac{4\pi R^{3}}{3} \bar{\rho}_{\rm pre}(<R) v.
\end{equation} 
Using $\bar{n}_{\rm H}^{\rm pre}(<R) = X \bar{\rho}_{\rm pre}(<R) / m_{\rm p}$ (where $X=0.75$ is the hydrogen mass fraction), we obtain
\begin{align}
\label{mean los density}
\bar{n}_{\rm H}^{\rm pre}(<R) & \approx  \frac{3 X f_{\rm w}}{4 \pi R^{2} c m_{\rm p}} \frac{L_{\rm bol}}{v^{2}} \\ \notag
& \approx 0.02~{\rm cm^{-3}} f_{\rm w}  
\left( \frac{L_{\rm bol}}{\rm 10^{47}~erg~s^{-1}} \right)
\left( \frac{R}{\rm 3~kpc} \right)^{-2} \\ \notag
&~~~~~~~~~~~~~~~~~~~~~ \times \left( \frac{v}{\rm 5,000~km~s^{-1}} \right)^{-2}.
\end{align} 
At such densities, the post-shock gas would have no chance to cool in a flow time, making it impossible to form FeLoBALs at large distances via interaction with ``average'' ambient conditions.

Because FeLoBALs are rare \citep[only $\sim 1/1,000$ of bright optical quasars show transitions suitable for photoionization analysis; e.g.,][]{2009ApJ...706..525M}, however, we may appeal to rare circumstances to produce them. 
Consider the situation illustrated in Figure \ref{cartoon}. 
A moderately dense clump, for example analogous to a cold neutral medium cloud, lies along the line of sight. 
Let $n_{\rm H}^{\rm c,i}$ be the initial density of the cloud and $a_{\rm i}$ its radius. 
The cloud is impacted by a blast wave from the quasar traveling at velocity $v_{\rm sh}\approx5,000$ km s$^{-1}$, as above. 
As the blast wave overtakes the cloud, a shock wave is driven into it at a velocity $v_{\rm sh,c} \approx v_{\rm sh} \sqrt{n_{\rm H}^{\rm pre} / n_{\rm H}^{\rm c,i}}$, set by the requirement that the post-shock gas in the cloud be in pressure equilibrium with the surrounding medium \citep[e.g.,][]{1975ApJ...195..715M}. 
The corresponding post-shock temperature is then $T_{\rm sh,c}=T_{\rm sh} ( n_{\rm H}^{\rm pre}/n_{\rm H}^{\rm c,i} )$, and we define a cloud crushing time as
\begin{equation}
t_{\rm cc} \equiv \frac{a_{i}}{v_{\rm sh,c}} \approx \frac{a_{i}}{v_{\rm sh}} \sqrt{ \frac{n_{\rm H}^{\rm c,i}}{n_{\rm H}^{\rm pre}} }. 
\end{equation}
Identifying the cloud size $\Delta R = 2 a_{\rm i}$ and $v=3v_{\rm sh}/4$, equation (\ref{KH time scale}) implies that the cloud is destroyed in $t_{\rm KH} = 8 \kappa t_{\rm cc}/ 3$.  

At the same time, ram pressure from the hot external medium accelerates the cloud, which becomes co-moving with the hot shocked gas on a drag time
\begin{equation}
t_{\rm drag} \approx \frac{4}{3} \frac{a_{\rm i}}{v} \frac{n_{\rm H}^{\rm c,i}}{4 n_{\rm H}^{\rm pre}},
\end{equation} 
i.e. the time necessary for the hot flow to sweep the cloud with a mass equal to its own. 
The requirement that the cloud be accelerated before it is torn apart is
\begin{equation}
\label{acc req}
\frac{t_{\rm drag}}{ t_{\rm KH}} \lesssim 1 \Leftrightarrow
\frac{n_{\rm H}^{\rm c,i}}{4 n_{\rm H}^{\rm pre}} \lesssim \left( \frac{3}{2} \kappa \right)^{2}. 
\end{equation}
We further require that the cooling time behind the shock driven into the cloud be less than a cloud crushing time, so that the shock is radiative and leaves cold material behind.
Such radiative cooling also prolongs the cloud survival with respect to the Kelvin-Helmholtz instability \citep[e.g.,][]{2002A&A...395L..13M, 2009ApJ...703..330C}, allowing values $\kappa \approx 10$. 
For $10^{5} \lesssim T \lesssim 4\times 10^{7}$ K, the isobaric cooling time for ionized gas is
\begin{equation}
t_{\rm cool} \approx 1.5\times10^{4}~{\rm yr}~\left( \frac{T_{\rm sh,c}}{\rm 10^{6}~K} \right)^{1.6} \left( \frac{4 n_{\rm H}^{\rm c,i}}{\rm 10~cm^{-3}} \right)^{-1},
\end{equation}
where we have adopted the approximation to the cooling function from \cite{1977ApJ...215..213M}, and the formula accounts for a factor of 4 compression through the strong shock. 
Combining the expressions above,
\begin{align}
\label{cooling req}
\frac{t_{\rm cool}}{t_{\rm cc}} & \approx 1 
\left( \frac{v_{\rm sh}}{\rm 5,000~km~s^{-1}} \right)^{4.2} 
\left( \frac{n_{\rm H}^{\rm pre}}{\rm 0.02~cm^{-3}} \right)^{2.1} \\ \notag
& \times \left( \frac{n_{\rm H}^{\rm c,i}}{\rm 18~cm^{-3}} \right)^{-2.1}
\left( \frac{N_{\rm H}^{\rm c,i}}{\rm 1.4\times10^{20}~cm^{-2}} \right)^{-1},
\end{align}
where $N_{\rm H}^{\rm c,i} \equiv 2 a_{\rm i} n_{\rm H}^{\rm c,i}$ is the column density of the original cloud. 
In this formula, the fiducial values for $n_{\rm H}^{\rm pre}$ and $n_{\rm H}^{\rm c,i}$ were chosen to satisfy $t_{\rm drag}\approx t_{\rm KH}$ with $\kappa=10$ (c.f., eq. \ref{acc req}), i.e. to ensure that the cloud is accelerated before being shredded. 
It follows that column densities $N_{\rm H}^{\rm c,i} \gtrsim10^{20}$ cm$^{-2}$ are necessary for the post-shock gas with $v_{\rm sh} \approx 5,000$ km s$^{-1}$ to cool and produce conditions suitable for FeLoBALs. 

For $t > t_{\rm drag},~t_{\rm KH}$, the main cloud has been shredded into multiple cloudlets traveling at $v \sim v_{\rm sh}$. 
Since $t_{\rm cool} < t_{\rm cc}$, these cloudlets cannot maintain pressure balance with the surrounding medium and are therefore compressed. 
These photoionized cloudlets cool to $T\approx10^{4}$ K, where the cooling function is strongly suppressed, and reach densities up to (in pressure equilibrium with the hot gas)
\begin{align}
\label{final density}
n_{\rm H}^{\rm c,f} & \approx 4 n_{\rm H}^{\rm pre} \left( \frac{T_{\rm sh}}{\rm 10^{4}~K} \right ) \\ \notag
& \approx 3,000~{\rm cm^{-3}} 
\left( \frac{n_{\rm H}^{\rm pre}}{\rm 0.02~cm^{-3}} \right)
\left( \frac{v_{\rm sh}}{\rm 5,000~km~s^{-1}} \right)^{2}.
\end{align}
The predicted densities, column densities,\footnote{In the radiative case, the original cloud does not undergo significant lateral expansion \citep[as it would adiabatically; e.g.][]{1994ApJ...420..213K}. At least before they mix significantly with the hot medium, the integrated column density of the cloudlets is therefore similar to that of the original cloud, provided that their collective covering factor is preserved during the shredding. We therefore identify $N_{\rm H}^{\rm c,i} \sim N_{\rm H}^{\rm BAL}$.} and temperatures of these cloudlets are in excellent agreement with the inferred properties of the FeLoBALs summarized in Table \ref{FeLoBAL properties table}. 

Furthermore, this picture naturally explains the observations of multiple absorption components along the line of sight with approximately the same distance from the SMBH \citep[e.g.,][]{2010ApJ...713...25B}, as some cloudlets may only be partially accelerated. 
The supra-thermal line widths measured (hundreds of km s$^{-1}$) could also result from a combination of blended components and velocity shear during the shredding process. 

In our model, the fact that FeLoBALs arise along particular sight lines with over-dense material is also in agreement with their host quasars being redder than average \citep[even among BAL quasars; e.g.,][]{1992ApJ...390...39S, 2000ApJ...538...72B, 2008ApJ...674...80U}.
It is also possible that dust is present in other clouds along the line of sight, which in general could be seen in absorption at different velocity offsets. 
However, since strong absorption at the galaxy systemic redshift is often not seen in FeLoBALs, this suggests that the dust responsible for the reddening is part of the FeLoBAL gas, and therefore that the dust survived the cloud crushing shocks. 
We note that the intra-cloud shocks have velocities much lower than the quasar blast wave owing to the cloud over-density. Even for a quasar blast wave with velocity $v_{\rm sh}=5,000$ km s$^{-1}$, for the maximum density contrast allowed by equation (\ref{acc req}) with $\kappa=10$, the cloud crushing shocks have velocities $v_{\rm sh,c} \sim 170$ km s$^{-1}$. 
At such velocities, it is reasonable for at least some dust to survive in the FeLoBAL gas \citep[e.g.,][]{1989IAUS..135..431M}. 
Indeed, although the dust is likely to be eventually destroyed with the cloud itself, the short lifetimes of order a few $1,000$ yr (\S \ref{created in situ}) imply that there is little time for dust sputtering processes to operate while the cloud is present.

\subsection{Selection Effects and Predicted Trends}
\label{selection effects}
In reality, the hot flow likely entrains a wide spectrum of clouds. 
In the scenario outlined above, selection effects explain the relatively narrow range of observed FeLoBAL properties. 

Absorption by low-ionization species such as FeII, HeI, and SiII useful for photoionization modeling selects ionization parameters $U_{\rm H}\sim10^{-2}-10^{-3}$, as well as temperatures sufficiently low to avoid collisional ionization of those species. 
As equations (\ref{acc req}) and (\ref{cooling req}) show, cool gas remains behind the blast wave \emph{and} is accelerated to a velocity comparable to the hot flow only for clump column densities 
\begin{equation}
\label{cooling and acc NH}
N_{\rm H}^{\rm c,i} \gtrsim 10^{20}~{\rm cm^{-2}} 
\left(
\frac{v_{\rm sh}}{\rm 5,000~km~s^{-1}}
\right)^{4.2}.
\end{equation} 
Interestingly, there is evidence that FeLoBALs located closer to SMBHs ($\sim 1-10$ pc vs. $\gtrsim 1$ kpc) have column densities higher by factors $\sim 10-100$ \citep[e.g.,][]{2010ApJ...709..611D}, a trend in agreement with our model if the blast wave decelerates with radius.\footnote{Since not all cloudlets need be fully accelerated to the hot flow velocity, their velocity may not always reflect the blast wave velocity. Furthermore, the criterion in equation (\ref{cooling and acc NH}) is strengthened for values of $n_{\rm H}^{\rm c,i}/n_{\rm H}^{\rm pre}$ smaller than fiducially assumed in equation (\ref{cooling req}). There is therefore not necessarily a well-defined relationship between $N_{\rm H}$ and $R$, or $N_{\rm H}$ and $v$.}

Our model also naturally explains why the FeLoBAL absorbers for the bright quasars in Table \ref{FeLoBAL properties table} are all located at relatively large radii $R\sim1-3$ kpc. Indeed, the observational selection of a narrow range of ionization parameter $U_{\rm H} \propto L_{\rm bol} / R^{2} n_{\rm H}$ defines a relationship between the quasar luminosity $L_{\rm bol}$ and the FeLoBAL radius $R$. Since the FeLoBAL gas density $n_{\rm H}$ is set by pressure equilibrium as in equation (\ref{final density}), this relationship in fact implies that the FeLoBAL absorbers in luminous quasars would be typically found at large radii.

\begin{table*}
\centering
\caption{Galaxy-Scale Mass Outflow Rates, Kinetic Luminosities, and Momentum Fluxes Inferred from the FeLoBALs in Table \ref{FeLoBAL properties table}\label{feedback table}}
\begin{tabular}{|cccccccc|}
\hline\hline
QSO                               & $\dot{M}_{\rm shell}(\Omega=0.2)$ & $\dot{E}_{\rm k}^{\rm shell}$ & $\dot{P}^{\rm shell}$ & $\dot{M}_{\rm hot}(\Omega_{\rm hot}=1)$ & $\dot{E}_{\rm k}^{\rm hot}$ & $\dot{P}^{\rm hot}$ & $\dot{E}_{\rm k}^{\rm QSO}$ \\
                                      & M$_{\odot}$ yr$^{-1}$ & $\%L_{\rm bol}$ & $L_{\rm bol}/c$ & M$_{\odot}$ yr$^{-1}$ & $\%L_{\rm bol}$ & $L_{\rm bol}/c$ & $\%L_{\rm bol}$ \\
\hline
SDSS J0838+2955 & $560$ & $1.2$ & 1.5 & 1,000 & 2.2 & 2.7 & 3.5 \\ 
SDSS J0318-0600 & $150$ & $0.2$ & 0.2 & 1,100 & 1.2 & 1.8 & 2.0 \\ 
QSO 2359-1241 & $36$ & $0.05$ & 0.2 & 2,400 & 3.1 & 13.4 & 5.0 \\ 
\hline
\tablecomments{
Values assuming that the outflow is confined to a cold thin shell (eq. \ref{Mdot shell}) are labeled `shell.' 
Values assuming that the FeLoBALs arise in radiative shocks (eq. \ref{Mdot hot}) and are entrained by an energetically-dominant hot flow are indicated by the label `hot.' 
For these, the FeLoBAL gas is assumed to be co-moving and in pressure equilibrium with the shocked ambient medium. 
A covering factor $\Omega=0.2$ is assumed for the cold thin shells for comparison with previous work, but a more physically plausible covering factor $\Omega_{\rm hot}=1$ is adopted for the hot gas at large radii (\S \ref{inferring feedback}).
The last column, $\dot{E}_{\rm k}^{\rm QSO}$, lists the kinetic luminosities of the small-scale quasar wind inferred assuming a self-similar adiabatic wind solution (see the Appendix). 
If the FeLoBAL gas is not fully accelerated to the velocity of the hot gas, $\dot{M}_{\rm hot}$ should be interpreted as an upper limit, and $\dot{E}_{\rm k}^{\rm hot}$ and $\dot{E}_{\rm k}^{\rm QSO}$ as lower limits. 
$\dot{E}_{\rm k}^{\rm QSO}$ provides a more complete estimate of the quasar wind energetics than $\dot{E}_{\rm k}^{\rm hot}$, which neglects the fraction of the energy thermalized in the hot flow. 
}
\end{tabular}
\end{table*}

\section{INFERRING THE FEEDBACK EFFICIENCY}
\label{inferring feedback}
The physical picture of FeLoBALs developed here allows us to relate their observed properties to those of the underlying quasar outflows.

Consider first the common assumption that the outflowing material is confined to a partial, cold, thin shell that has a covering factor $\Omega$ and a column density equal to that of the FeLoBAL, $N_{\rm H}^{\rm BAL}$  \citep[e.g.,][]{2010IAUS..267..350A}. 
Since the mass of the shell $M_{\rm shell} = 4 \pi R^{2} N_{\rm H}^{\rm BAL} m_{\rm p} \Omega / X$, the mass outflow rate is given by
\begin{equation}
\label{Mdot shell}
\dot{M}_{\rm shell} = \frac{8 \pi R N_{\rm H}^{\rm BAL} m_{\rm p} v} {X} \Omega,
\end{equation}
where $v = \dot{R}$. 
The corresponding kinetic luminosity and momentum flux are then 
\begin{equation}
\label{Edot k}
\dot{E}_{\rm k} = \frac{1}{2} \dot{M}_{\rm shell} v^{2}
\end{equation}
and
\begin{equation}
\label{Pdot}
\dot{P} = \dot{M}_{\rm shell} v.
\end{equation} 
Motivated by the observed incidence of general (mostly high ionization) broad absorption lines in quasars, these expressions have usually been applied with a covering factor $\Omega \approx 0.2$ \citep[e.g.,][]{2009ApJ...706..525M, 2010ApJ...709..611D}. 
We return to this point below, where we argue that this may generally not be a valid assumption for FeLoBALs.

In our picture, FeLoBALs are rare tracers of an underlying shock-heated medium, fortuitously encountered along the line of sight.  A more appropriate estimate of the mechanical properties of the outflow can be obtained by replacing the column density of the FeLoBAL by that of the hot gas in equation (\ref{Mdot shell}), and similarly for the velocity and covering factor, i.e.
\begin{equation}
\label{Mdot hot}
\dot{M}_{\rm hot} = \frac{8 \pi R N_{\rm H}^{\rm hot} m_{\rm p} v_{\rm hot}} {X} \Omega_{\rm hot}. 
\end{equation}
The shell approximation remains reasonable, since at large radii most of the hot gas is swept up, shocked ambient ISM. 

The properties of the hot flow required in equation \ref{Mdot hot} are not directly measured, but can be inferred by making assumptions motivated by our model.   One subtlety is that in a realistic outflow, the ambient medium is separated from the direct quasar wind by a forward shock, a contact discontinuity, as well as a reverse shock.   We discuss this in detail in the Appendix but it does not change the conclusions drawn here using simpler arguments. 

We begin by assuming that the FeLoBAL gas is fully accelerated by the hot gas, $v \approx v_{\rm hot}$, and in pressure equilibrium with it:
\begin{equation}
n_{\rm H}^{\rm hot} \approx n_{\rm H}^{\rm BAL} \left( \frac{\rm 10^{4}~K}{T_{\rm hot}} \right). \label{eq:nhot}
\end{equation}
As before, $T_{\rm hot} \approx T_{\rm sh}(v_{\rm sh}=4v/3)$, where $T_{\rm sh}$ is given by equation (\ref{Tsh}).    The column density of the hot gas can be estimated as
\begin{equation}
\label{NH hot}
N_{\rm H}^{\rm hot} \approx \bar{n}_{\rm H}^{\rm pre}(<R) R.
\end{equation}
If the FeLoBAL is located in the shocked ambient medium and close to the forward shock (see the Appendix), then $n_{\rm H}^{\rm hot} \approx 4 n_{\rm H}^{\rm pre}$. 
Under these assumptions,
\begin{equation}
N_{\rm H}^{\rm hot} \approx \frac{1}{4} N_{\rm H}^{\rm BAL} \left( \frac{R}{\Delta R} \right) \left( \frac{\rm 10^{4}~K}{T_{\rm hot}} \right),
\end{equation}
where  $\Delta R$ is the thickness of the FeLoBAL absorber (as before). 
Written this way, all the quantities necessary to evaluate equation (\ref{Mdot hot}) are provided by photoionization modeling of the FeLoBAL (e.g., Table \ref{FeLoBAL properties table}), except $T_{\rm hot}$ which is derived from $v$, and $\Omega_{\rm hot}$.   The  energy and momentum fluxes of the outflow follow straightforwardly given $\dot M_{\rm hot}$ and $v_{\rm hot}$.

It is important to stress that $\dot M_{\rm hot}$ in equation \ref{Mdot hot} is not the intrinsic mass outflow rate in the quasar wind.  Instead, $\dot M_{\rm hot}$  is the rate at 
which the ambient ISM material is being swept up by the quasar outflow, i.e., it is the galaxy-scale outflow rate of shocked gas.   In the model of FeLoBALs advocated here, the observations cannot directly constrain the intrinsic quasar mass outflow rate because most of the mass that is pressure confining the BAL gas is likely swept-up ISM material, rather than the direct quasar wind.   However, the observations do directly constrain the galaxy-scale outflow rate $\dot M_{\rm hot}$ and the corresponding energy and momentum fluxes $\dot E_{\rm k}^{\rm hot}$ and $\dot P^{\rm hot}$. 
Note that $\dot E_{\rm k}^{\rm hot}$ corresponds only to the bulk kinetic energy of the hot gas, neglecting a fraction $\sim 1$ that is instead thermalized. 
The derivation of the small-scale quasar wind kinetic luminosity $\dot{E}_{\rm k}^{\rm QSO}$ for a self-consistent outflow solution in the Appendix, on the other hand, accounts for all the energy. 
Accordingly, $\dot{E}_{\rm k}^{\rm QSO} \sim 2 \dot{E}_{\rm k}^{\rm hot}$.

In Table \ref{feedback table}, we give the galaxy-scale mass outflow rates, kinetic luminosities, and momentum fluxes derived from the FeLoBAL systems in Table \ref{FeLoBAL properties table}, evaluated using both the cold thin shell approximation (eq. \ref{Mdot shell}), and assuming that the FeLoBAL is only a tracer of the underlying shocked medium as above (eq. \ref{Mdot hot}). 
For the former case, we fiducially adopt $\Omega=0.2$ for comparison with previous work. 
For the latter, we instead fiducially assume $\Omega_{\rm hot}=1$. 
A covering factor $\Omega_{\rm hot}\approx 1$ seems more physically plausible at kilo-parsec radii, because in the absence of significant obstacles the hot bubble will tend to become more spherical as it expands. 
This applies even if the energy is, for example, injected in the form of high-ionization BAL systems with $\Omega \sim 0.2$ from an accretion disc wind at $R<1$ pc. 
Momentum conservation actually suggests that the outflows traced by the FeLoBALs in Table \ref{FeLoBAL properties table} are propagating along sight lines of low gas density, consistent with a spherically symmetric geometry and possibly owing to an earlier phase of feedback which cleared most of the gas out. 
A dense galactic disc might impede the outflow along certain directions, but should have a minor effect provided it is thin.    
We however cannot exclude the possibility of viewing the quasars along clear sight lines by chance. 
Thus, there is significant uncertainty in the normalization of the energetics in Table \ref{feedback table} due to the covering factor assumption. 

As Table \ref{feedback table} shows, the mass outflow rates inferred for the hot flows (and consequently $\dot{E}_{\rm k}$ and $\dot{P}$) are systematically higher than those implied for the cold thin shells, even after normalizing to the same covering factor.\footnote{This is violated by a factor $<3$ for SDSS J0838+2955, but this is only marginally significant at the level of our estimates. It is also possible that the absorber is not fully accelerated.}  
Indeed, this must generically be the case for fully accelerated FeLoBALs, since then
\begin{equation}
\label{Mdot shell vs Mdot hot}
\frac{\dot{M}_{\rm shell}}{\dot{M}_{\rm hot}} = \frac{N_{\rm H}^{\rm BAL}}{N_{\rm H}^{\rm hot}} \left( \frac{\Omega}{\Omega_{\rm hot}} \right),
\end{equation}
and full acceleration requires $N_{\rm H}^{\rm hot} \gtrsim N_{\rm H}^{\rm BAL}$ (i.e., that the cool cloud be swept with a mass equal to its own). 
Another bound on the ratio of the column densities of the cold and hot gas can be derived in the context of our model by combining the momentum conservation and cooling conditions (eq. \ref{mean los density} and \ref{cooling and acc NH}):
\begin{align}
\label{NH BAL vs NH hot}
\frac{N_{\rm H}^{\rm BAL}}{N_{\rm H}^{\rm hot}} \gtrsim \frac{1}{f_{\rm w}} 
\left( \frac{L_{\rm bol}}{\rm 10^{47}~erg~s^{-1}} \right)^{-1}
& \left( \frac{R}{\rm 3~kpc} \right) \\ \notag
& \times
\left( \frac{v}{\rm 5,000~km~s^{-1}} \right)^{6.2}.
\end{align}
This expression is, again, in good agreement with the results in Table \ref{feedback table}. 
Indeed, for SDSS J0838+2955 ($v=4,900$ km s$^{-1}$) $\dot{M}_{\rm shell}\sim \dot{M}_{\rm hot}$ for the same covering factor, as combining equations (\ref{Mdot shell vs Mdot hot}) and (\ref{NH BAL vs NH hot}) indicates should be the case. 
However, the steep dependence of equation (\ref{NH BAL vs NH hot}) on $v$ implies that $\dot{M}_{\rm shell}$ for QSO 2359-1241 ($v=1,400$ km s$^{-1}$) \emph{can be} much less than $\dot{M}_{\rm hot}$, and in fact it is found to be.

Although we have so far assumed that the FeLoBAL gas is fully accelerated, i.e. co-moving with the hot flow, this may not always hold. 
Provided only that the cool absorbers are pressure confined, $\dot{M}_{\rm hot} \propto v_{\rm hot}^{-1}$, $\dot{E}_{\rm k}^{\rm hot} \propto v_{\rm hot}$, and $\dot{P}^{\rm hot}$ is independent of $v_{\rm hot}$. 
In particular, a useful lower limit on the kinetic luminosity of the outflow is obtained (for a given $\Omega_{\rm hot}$) if the FeLoBAL velocity $v \leq v_{\rm hot}$, while the derived $\dot{M}_{\rm hot}$ is then a lower limit. 
The values in Table \ref{feedback table} should more generally be interpreted this way. 

FeLoBALs with properties identical to those in our blast wave model can also be produced in a smooth, supersonic wind (with a density profile $\propto R^{-2}$ and velocity $v_{\rm in}$) entraining an interstellar cloud, since the cloud crushing then proceeds similarly \citep[e.g.,][]{1994ApJ...420..213K}. 
In this case, the cold gas reaches a terminal velocity $v$ satisfying $N_{\rm H}^{\rm w}/N_{\rm H}^{\rm BAL} \approx (v/v_{\rm in})^{2}<1$, where $N_{\rm H}^{\rm w}$ is the column density through the wind at radius $R$. 
This ``pure wind'' scenario is less physically plausible since FeLoBALs have velocity offsets $\sim 1,000-5,000$ km s$^{-1}$, compared to accretion disc wind velocities $\sim0.1c=30,000$ km s$^{-1}$.   This suggests that the quasar wind has decelerated significantly along the line of sight by sweeping up  ambient ISM. 

\section{DISCUSSION}
\label{discussion}

We have argued that the FeLoBAL absorption systems seen in a small fraction of quasars are physically distinct from the larger class of broad line quasar absorbers. 
The majority of high-ionization BALs are located at radii $R<1$ pc from the accreting supermassive black hole and are likely directly accelerated by it, possibly in an accretion disc wind \citep[e.g.,][]{1995ApJ...451..498M}.  The kinematics of most LoBAL quasars also appear to differ from those of FeLoBALs: \citet[]{1993ApJ...413...95V} stress that in the LoBALs they study (which do not show FeII absorption) the Mg~II and Al~III features are narrower (by factors of several in some cases) than the C~IV troughs, and the Mg~II and Al~III features tend to lie at the low-velocity ends of the C~IV trough. This contrasts with the FeLoBALs discussed by \citet[]{2009ApJ...706..525M}, \citet[]{2010ApJ...709..611D} and \citet{2010ApJ...713...25B}, where the C IV troughs have similar total velocity width to that of Mg II and Al III, as well as similar velocity structure, in which the individual features of different ions line up in velocity space. The latter behavior strongly suggests that the C IV absorbing gas is co-spatial with the Mg II and Al III gas in FeLoBALs, unlike the C~IV gas in the \citet[]{1993ApJ...413...95V} objects.

Consistent with our argument that the FeLoBAL systems are physically distinct from other BAL systems, we have shown that the FeLoBALs form \emph{in situ} in the ISM of the host galaxy. Indeed, when a quasar blast wave impacts a sufficiently dense clump along the line of sight, the cloud crushing shocks are radiative and leave cool material behind. 
We derived the conditions under which (1) the post-shock gas cools and (2) the cool gas is accelerated to a velocity $\gtrsim1,000$ km s$^{-1}$:  the required properties are in excellent agreement with those inferred for FeLoBALs. 

This new physical picture of FeLoBALs has important implications for measurements of the efficiency with which quasars convert their luminosity into mechanical feedback on galactic scales. 
Unlike the cold thin shell approximation often assumed \citep[e.g.,][]{2010IAUS..267..350A}, in our picture FeLoBALs are only rare tracers of an underlying outflow.   In particular, the observed FeLoBAL properties constrain the properties of the ambient ISM shock heated by the quasar blast wave.  We find that the mass outflow rate, kinetic luminosity, and momentum flux of the hot flow can be larger, sometimes by an order of magnitude or more, than those previously estimated for the cold gas only. 
In Table \ref{feedback table}, we give the values derived for three bright quasars in the literature, using both the cold thin shell approximation and in the context of our radiative shock model. 
In our model, $\dot{E}_{\rm k}^{\rm QSO}\approx2-5$\% $L_{\rm bol}$, with corresponding momentum fluxes $\dot{P}^{\rm hot} \approx 2-15 L_{\rm bol}/c$. In the case of QSO 2359-1241, the kinetic luminosity $\dot{E}_{\rm k}^{\rm QSO}\approx5$\% $L_{\rm bol}$ is $\sim 100 \times$ larger than implied by the cold thin shell approximation.

Our inferred values for the kinetic power in quasar outflows on galactic scales are comparable to those assumed in models successful in reproducing the $M-\sigma$ relation \citep[e.g.,][]{2003ApJ...595..614W, 2005Natur.433..604D}.    Moreover, for a given covering factor of hot gas $\Omega_{\rm hot}\approx1$, our inference of $\dot E_k^{\rm QSO}$ is a lower limit if the cold BAL gas is not fully accelerated to the velocity of the hot gas. 
The momentum fluxes we find for FeLoBALs in Table \ref{feedback table} are also comparable to the values $\dot{P} \approx 5-10 \, L_{\rm bol}/c$ that appear sufficient to drive powerful galaxy-wide outflows in simulations that model black hole feedback by injecting quasar winds into the surrounding ISM \citep[][]{debuhr11}.

The mass outflow rates and energy and momentum fluxes inferred in Table \ref{FeLoBAL properties table} apply on the galactic scales where the absorbers are detected ($R \sim$ kpc).
These need not correspond to the small-scale mechanical properties of the quasar wind. 
In particular, galactic-scale mass outflow rates are in general higher than those of the direct quasar wind because of swept up material:  the large values in Table \ref{FeLoBAL properties table} therefore need not imply super-Eddington accretion disc winds.

The FeLoBALs in the three bright quasars studied here may trace a particular stage of feedback. 
In fact, the observations of cool absorbers moving at $v\sim 5,000$ km s$^{-1}$ at $R\sim 3$ kpc from the black hole imply relatively clear sight lines. Otherwise, the quasar wind would have been  decelerated more significantly by sweeping up ambient ISM. 
One possibility is that we are witnessing late accretion events, with the outflows propagating into a tenuous ISM that has already been mostly evacuated in an earlier phase. 
The powerful molecular outflows recently detected in local ultra-luminous infrared galaxies (ULIRGs) may provide evidence of such gas clearing \citep[e.g.,][]{2010A&A...518L.155F, 2010A&A...518L..41F, 2011ApJ...733L..16S}, though star formation may also contribute to driving these winds \citep{chung11}.  Intriguingly, the kinetic luminosities and momentum fluxes of the ULIRG molecular outflows are similar to those we infer for the FeLoBAL quasar outflows, suggesting that AGN are indeed capable of driving the ULIRG outflows. 

Independent constraints on the evolutionary stage of FeLoBAL quasars, for instance from their broad band spectral energy distributions, would be valuable in distinguishing the late feedback possibility from that of a chance geometrical effect. 
Subject to the significant limitations of small and heterogeneous samples, recent observations appear to support the hypothesis that at least some FeLoBAL quasars are observed toward the end of their starburst and in the last stages of blowing out their surrounding ISM \citep[][]{2008ApJ...674...80U, 2007ApJ...667..149F, 2010ApJ...717..868F}. 
Since much of the black hole growth is predicted to occur while the AGN is obscured, appearing in the optical only after it has blown out the obscuring material \citep[e.g.,][]{2005ApJ...625L..71H}, this late stage of feedback picture is consistent with the ``youthful'' quasar hypothesis for FeLoBALs \citep[][]{2002ApJS..141..267H}. 

\section*{Acknowledgments}
CAFG is supported by a fellowship from the Miller Institute for Basic Research in Science and NASA grant 10-ATP10-0187. 
EQ is supported in part by the David and Lucile Packard Foundation.   NM is supported in part by the Canada Research Chair program and by NSERC of Canada. 

\appendix
\section{RELATIONSHIP BETWEEN THE SHOCKED AMBIENT MEDIUM AND THE QUASAR WIND}
\label{append}
\begin{figure}
\begin{center} 
\includegraphics[width=0.45\textwidth]{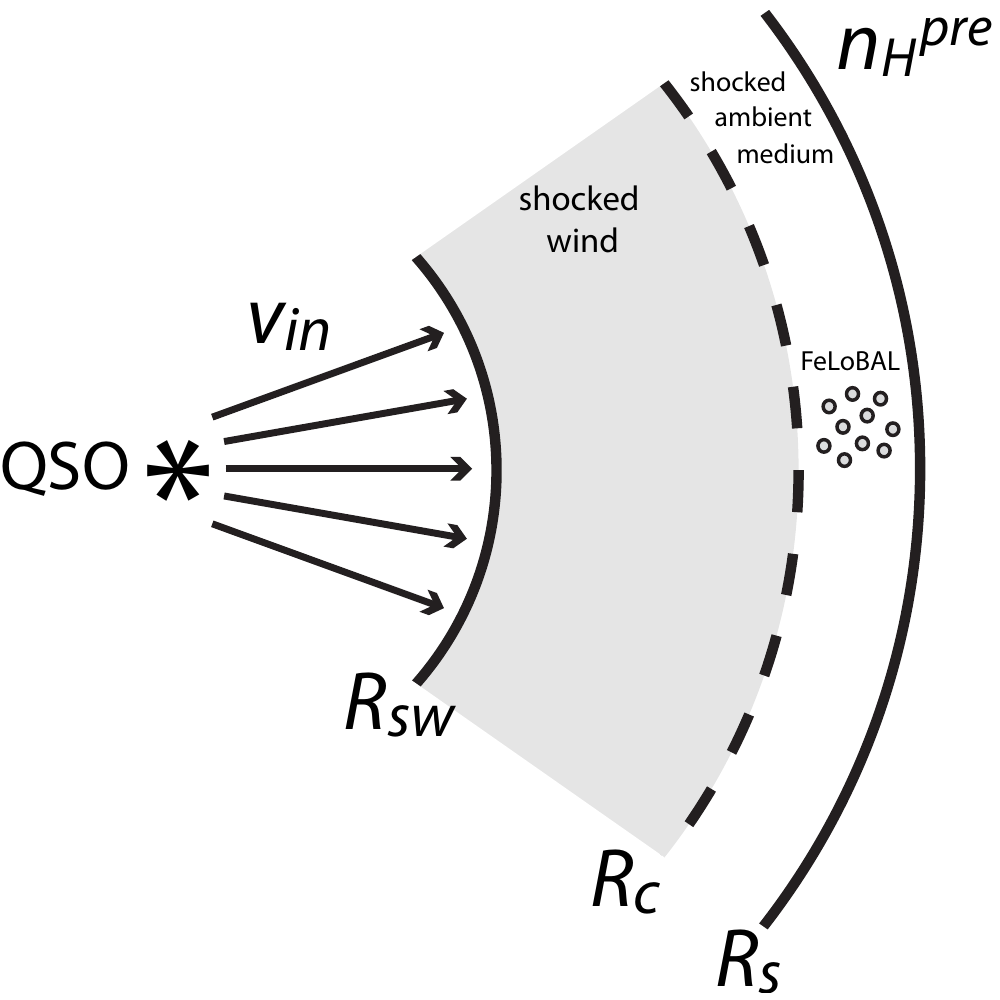}
\end{center}
\caption[]{Quasar outflow structure. The shocked ambient medium (bounded by $R_{\rm s}$), in which the FeLoBALs are envisioned to form, is separated from the direct quasar wind by a reverse shock at $R_{\rm sw}$ and a contact discontinuity at $R_{\rm c}$.}
\label{wind details} 
\end{figure}

In \S \ref{inferring feedback}, we inferred the mechanical properties of the quasar outflow based on those of the hot, shocked ambient medium at $R$. 
In realistic outflow solutions (see Fig. \ref{wind details}), the quasar wind is separated from that medium by a reverse shock at $R_{\rm sw}$ and a contact discontinuity at $R_{\rm c}$. We denote the radius of the forward shock by $R_{\rm s}$. 

An important question is how $\dot{E}_{\rm k}^{\rm hot}$ relates to the kinetic luminosity of the wind in the immediate vicinity of the quasar, $\dot{E}_{\rm k}^{\rm QSO}$. 
For some purposes, such as comparing with galaxy simulations that inject energy near the BH \citep[e.g.,][]{2005Natur.433..604D}, this quantity is more directly relevant than the hot flow on galactic scales. 
From FeLoBALs at kilo-parsec radii, this can only be done assuming a global outflow solution. 

As we showed (eq. \ref{mean los density}), both the reverse and forward shocks are likely to be adiabatic for the FeLoBALs in Table \ref{FeLoBAL properties table}. 
For the case of a quasar wind with constant $\dot{E}_{\rm k}^{\rm QSO} = \frac{1}{2} \dot{M}_{\rm in} v_{\rm in}^{2}$ (where the subscript `in' denotes the initial properties of the wind) propagating in an uniform ambient medium of density $\rho_{\rm pre}$, \cite{1992ApJ...388..103K} showed that
\begin{equation}
\label{Rs}
R_{\rm s} = 0.884 \left( \frac{\dot{E}_{\rm k}^{\rm QSO}}{\rho_{\rm pre}} \right)^{1/5} t^{3/5}
\end{equation}
in this regime. 
This solution incorporates the shock jump conditions and the pressure balance requirement at the contact discontinuity in the dimensionless pre-factor. 
It also assumes that the outflow has reached radii sufficiently large that it is self-similar and no longer depends on the details of the energy injection. 

Taking the derivative, time can be eliminated from equation (\ref{Rs}) and it is straightforward to show that
\begin{equation}
\label{Lin}
\dot{E}_{\rm k}^{\rm QSO} \approx 8.6 \rho_{\rm pre} R_{\rm s}^{2} \dot{R}_{\rm s}^{3}.
\end{equation}
Since the FeLoBALs are formed upon impact with the forward shock in our model, and $t_{\rm evap} \ll t_{\rm flow}$ for the systems in Table \ref{FeLoBAL properties table}, we can usually assume $R\approx R_{\rm s}$. 
In general, $v \leq 3\dot{R}_{\rm s}/4$ and if the absorber is fully accelerated, then $v \approx 3\dot{R}_{\rm s}/4$. 
To estimate the density of the pre-shock ambient medium, it is necessary to assume that the FeLoBAL gas is in pressure equilibrium with the gas behind the forward shock:
\begin{equation}
\rho_{\rm pre} \approx \frac{n_{\rm H}^{\rm BAL} m_{\rm p}}{4 X} \left( \frac{\rm 10^{4}~K}{T_{\rm sh}(v_{\rm sh}=4v/3)} \right). 
\end{equation}

Under these assumptions, which are essentially the same that were necessary to estimate the properties of the hot flow in \S \ref{inferring feedback}, 
\begin{align}
\dot{E}_{\rm k}^{\rm QSO} \lesssim 2\times10^{46}~{\rm erg~s^{-1}}
& \left(
\frac{n_{\rm H}^{\rm BAL}}{\rm 10^{4}~cm^{-3}}
\right)
\left(
\frac{R}{\rm 3~kpc}
\right)^{2} \\ \notag
& \times 
\left(
\frac{v}{\rm 5,000~km~s^{-1}}
\right),
\end{align}
with equality holding in the case of full acceleration. 
This can be compared to the kinetic luminosity of the hot flow estimated using equation (\ref{Mdot hot}):
\begin{equation}
\label{Edotk hot vs Edotk QSO}
\frac{\dot{E}_{\rm k}^{\rm hot}}{\dot{E}_{\rm k}^{\rm QSO}} \approx 0.6 \Omega_{\rm hot},
\end{equation}
a constant $\sim 1$. 
This near equality is not surprising since we assume that the outflow is adiabatic, so that the energy in the flow is preserved with radius.
It does however formally confirm that the kinetic energy of the quasar wind can be estimated to a good approximation based on the properties of the hot flow probed by FeLoBAL observations. 
An order unity difference between $\dot{E}_{\rm k}^{\rm hot}$ and $\dot{E}_{\rm k}^{\rm QSO}$ is expected, as the derivation of $\dot{E}_{\rm k}^{\rm hot}$ ignores the fraction of the quasar wind kinetic luminosity that is converted into thermal energy, rather than bulk motion (\S \ref{inferring feedback}).  

Since the self-similar solution in equation (\ref{Rs}) depends only on $\dot{E}^{\rm QSO}_{\rm k}$ and $\rho_{\rm pre}$, it is however not possible to separately infer $\dot{M}_{\rm in}$ or $v_{\rm in}$ from the large-scale properties of the outflow.

\bibliography{references} 
 
\end{document}